\documentclass[preprint]{aastex62}  

\begin{document}

                           % The preamble begins here.
\title{A New High Perihelion Inner Oort Cloud Object: 2015 TG387}
%\correspondingauthor{Scott S. Sheppard}
%\email{ssheppard@carnegiescience.edu}

\author{Scott S. Sheppard}
\affil{Department of Terrestrial Magnetism, Carnegie Institution for Science, 5241 Broad Branch Rd. NW, Washington, DC 20015, USA, ssheppard@carnegiescience.edu}

\author{Chadwick A. Trujillo}
\affil{Northern Arizona University, Flagstaff, AZ 86011, USA}

\author{David J. Tholen}
\affil{Institute for Astronomy, University of Hawai'i, Honolulu, HI 96822, USA}

\author{Nathan Kaib}
\affil{HL Dodge Department of Physics and Astronomy, University of Oklahoma, Norman, OK 73019, USA}

\begin{abstract}  % Produces abstract

Inner Oort Cloud objects (IOCs) are Trans-Plutonian for their entire
orbits.  They are beyond the strong gravitational influences of the
known planets yet close enough to the Sun that outside forces are
minimal.  Here we report the discovery of the third known IOC after
Sedna and 2012 VP113, called 2015 TG387.  2015 TG387 has a perihelion
of $65 \pm 1$ au and semi-major axis of $1170 \pm 70$ au.  The
longitude of perihelion angle, $\bar{\omega}$, for 2015 TG387 is
between that of Sedna and 2012 VP113, and thus similar to the main
group of clustered extreme trans-Neptunian objects (ETNOs), which may
be shepherded into similar orbital angles by an unknown massive
distant planet, called Planet X or Planet Nine.  2015 TG387's orbit is
stable over the age of the solar system from the known planets and
Galactic tide.  When including outside stellar encounters over 4 Gyrs,
2015 TG387's orbit is usually stable, but its dynamical evolution
depends on the stellar encounter scenarios used.  Surprisingly, when
including a massive Planet X beyond a few hundred au on an eccentric
orbit that is anti-aligned in longitude of perihelion with most of the
known ETNOs, we find 2015 TG387 is typically stable for Planet X
orbits that render the other ETNOs stable as well.  Notably, 2015
TG387's argument of perihelion is constrained and its longitude of
perihelion librates about 180 degs from Planet X's longitude of
perihelion, keeping 2015 TG387 anti-aligned with Planet X over the age
of the solar system.  We find a power law slope near 3 for the
semi-major axis distribution of IOCs, meaning there are many more high
than low semi-major axis IOCs. There are about 2 million IOCs larger
than 40 km, giving a mass of $10^{22}$ kg. The IOCs inclination
distribution is similar to the scattered disk, with an average
inclination of 19 degs.

\end{abstract}


\section{Introduction}

Extreme Trans-Neptunian objects (ETNOs) have perihelia well beyond
Neptune and large semi-major axes ($a>150-250$ au).  The ETNOs have
only minimal interactions with the known giant planets and thus are
strongly sensitive to gravitational forces hundreds to thousands of au
from the Sun.  Thus, the ETNOs can be used to probe the solar system
beyond the Kuiper Belt (Trujillo and Sheppard 2014).

\begin{figure}
\epsscale{0.4}
\centerline{\includegraphics[angle=90,totalheight=0.4\textheight]{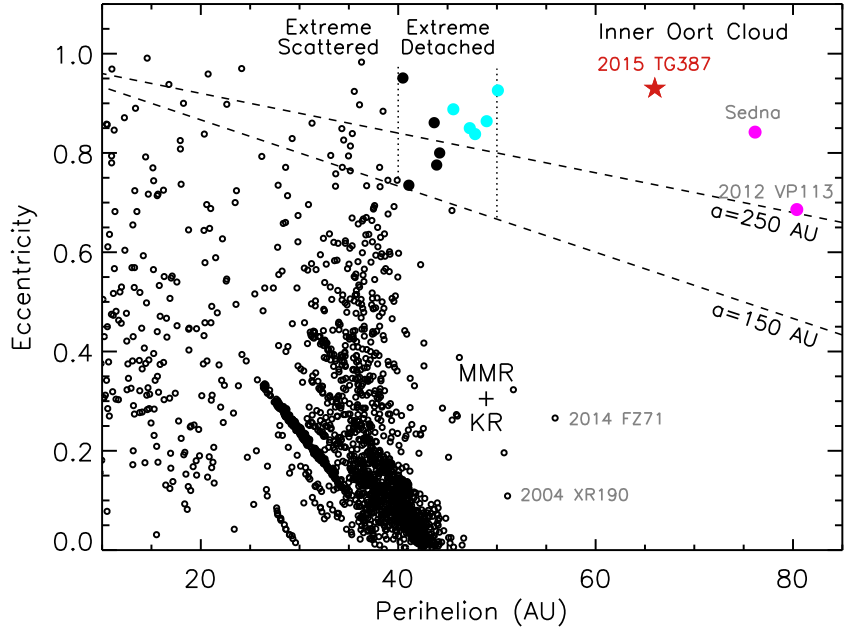}}
\caption{The perihelion versus eccentricity of small outer solar
  system objects with well known orbits as of July 2018 from the Minor
  Planet Center.  The red star shows our new Trans-Plutonian or Inner
  Oort Cloud object 2015 TG387 with a perihelion near 65 au. Objects
  above the 150-250 au semi-major axis dashed lines are considered
  extreme.  Bonafide Inner Oort Cloud objects are considered to have
  perihelion above 50-60 au (purple filled circles).  Extreme detached
  objects are mostly decoupled from the giant planets and have
  perihelion between about 40-45 and 50-60 au (strictest definition
  with $45<q<50$ au and $a>250$ au shown with filled blue circles,
  less extreme detached ETNOs shown with filled black circles). The
  less extreme detached ETNOs may have significant interactions with
  Neptune, while some may have a similar origin as the IOC
  objects. Extreme scattered disk objects have perihelia below 40 au
  and can have significant interactions with Neptune.  Outer Oort
  cloud objects have aphelion above a few thousand au and can have
  significant interactions with outside forces such as the Galactic
  and stellar tides.  Objects with relatively high perihelion beyond
  the Kuiper Belt edge at 50 au but only moderate eccentricity are
  likely created by a combination of past Neptune Mean Motion
  Resonances (MMR) and the Kozai Resonance (KR) and are detailed in
  Sheppard et al. (2016) and Kaib and Sheppard (2016).  All object
  uncertainties are generally smaller than the symbols.
\label{fig:kboeq2018paper} }
\end{figure}

The ETNOs can be separated into three sub-classes
(Figure~\ref{fig:kboeq2018paper}).  The scattered ETNOs have perihelia
below 38-45 au and likely were created from gravitational scattering
with Neptune and still have strong to moderate interactions with the
known giant planets (Brasser \& Schwamb 2015).  The detached ETNOs
have more distant perihelia of between about 40-45 to 50-60 au, but
could still have significant interactions with the known giant planets
(Gladman et al. 2002; Bannister et al. 2017).  Inner Oort Cloud
objects (IOCs) or Trans-Plutonian objects (TPOs) have perihelia
greater than 50-60 au and are too far from the giant planets to be
strongly influenced by them (Gomes et al. 2008).  The origins of the
eccentric IOC orbits likely require mechanisms that operated more
efficiently in the past such as stronger outside stellar tide forces
or uncatalogued forces in the outer solar system (Fernandez 1997;
Brown et al. 2004; Kenyon \& Bromley 2004; Madigan et al. 2018;
Sefilian \& Touma 2018).  The orbits of the IOCs thus inform us how
the distant solar system formed and currently interacts with its
surroundings.  The detached ETNOs may have evolved in a similar manner
as the IOCs or could be more similar to the scattered ETNOs.  Any
object with an aphelion beyond a few thousand au is considered an
outer Oort cloud object as outside forces such as the Galactic tide
and passing stars become strongly significant at these distances (Kaib
et al. 2009).

Sedna's large semi-major axis means that past stronger stellar tidal
forces experienced by the solar system in the Sun's birth cluster
would likely create Sedna's detached orbit, making this the preferred
formation mechanism of the IOCs (Kaib \& Quinn 2008; Brasser et
al. 2012).  2012 VP113's much smaller semi-major axis and even higher
perihelion compared to Sedna means it is harder for the stellar
cluster tidal formation scenario to work in creating 2012 VP113's
detached orbit (Trujillo and Sheppard 2014).

Trujillo and Sheppard (2014) noticed the ETNOs appear to have orbital
clustering in their argument of perihelia and were possibly asymmetric
in longitude and suggested there is a Super-Earth or larger mass
planet beyond a few hundred au shepherding these objects into similar
types of orbits.  Batygin and Brown (2016a) determined a possible
orbit a planet could have to cause the ETNOs to be aligned, with the
planet orbit needing to be eccentric, inclined and at several hundred
au.  After the above works, there have been several more in depth
analyses of the ETNOs and how they would evolve and interact under
such a massive planet in the outer solar system (Batygin \& Morbidelli
2017; Shankman et al. 2017a; Nesvorny et al. 2017; Iorio 2017; Khain
et al. 2018; Hadden et al. 2018; Li et al. 2018).  High inclination
trans-Neptuninan objects could be created by the planet (Batygin \&
Brown 2016b).  The planet should also cause resonant behavior in the
ETNOs (Malhotra et al. 2016; Millholland \& Laughlin 2017; Bailey,
Brown \& Batygin 2018), though the ETNOs may jump between various
orbital resonances and configurations (Becker et al. 2017).  One of
the most interesting behaviors of the ETNOs is the libration of the
longitude of perihelion with that of the unknown planet, which was
seen in numerical simulations of the ETNO 2013 FT28 by Sheppard and
Trujillo (2016).  2013 FT28 was announced after the possible planet
orbital parameters were reported by Batygin and Brown (2016a), so 2013
FT28 makes the case stronger that the massive unknown planet beyond a
few hundred au exists.

Here we detail the discovery of only the third object with a known
perihelion beyond 60 au, 2015 TG387, and how its orbit compares to the
other known IOCs and ETNOs.  We then discuss the stability of 2015
TG387 through numerical simulations we performed involving the known
major planets, Galactic tide, stellar passages and a possible massive
Planet X beyond a few hundred au.

\section{Basic Survey and Observation Details}

2015 TG387 was discovered in our ongoing survey for objects beyond the
Kuiper Belt edge.  This survey is discussed in detail in Trujillo and
Sheppard (2014) and Sheppard and Trujillo (2016), where several
extreme trans-Neptunian objects were discovered including 2012 VP113,
2014 SR349 and 2013 FT28. Here we report additional fields from the
survey (Table 1) and discuss the discovery and implications of 2015 TG387
within these fields, a more detailed paper on the full survey and the
discoveries made will follow later.

\begin{figure}
\epsscale{0.4}
\centerline{\includegraphics[angle=0,totalheight=0.4\textheight]{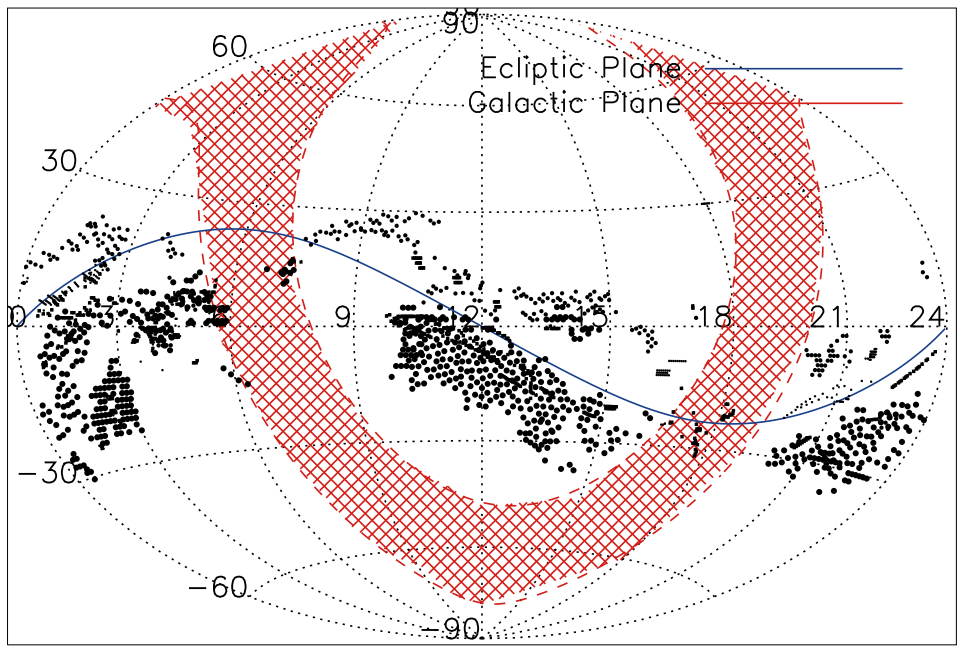}}
\caption{The newest survey fields were obtained using the HyperSuprime
  Camera on Subaru (medium circles) and the Dark Energy Camera on the
  CTIO 4m (large circles) as shown in Table 1.  Also included in the
  figure are the fields detailed in Sheppard and Trujillo (2016).  The
  Galactic plane is in red to $\pm15$ degrees of the center.  We
  covered about 1050 square degrees since Sheppard and Trujillo
  (2016), giving 2130 square degrees for the total survey to date.
\label{fig:MapDECAM2018paper} }
\end{figure}

For discovery, our survey mainly uses the 8.2 meter Subaru telescope
(atop Mauna Kea, Hawaii) in the Northern hemisphere with the 1.5
square degree HuperSuprime Camera (HSC) and the 4 meter Blanco
telescope (at Cerro Tololo Interamerican Observatory) in the Southern
hemisphere with the 2.7 square degree Dark Energy Camera (DECam)
(Flaugher et al. 2015).  Any objects found beyond about 50 au are
recovered months and years later with the 6.5 meter Magellan and 4
meter Discovery Channel Telescope (DCT) to obtain the orbits of the
objects.  The r-band Subaru HSC images generally have exposure times
of about 300 seconds, but exposure times are increased or decreased in
order to reach about 25.5 magnitudes in the r-band, depending on the
observing conditions for each night.  The DECam images use a very wide
broad band VR filter and are generally 420 seconds with exposure times
varied depending on the observing conditions to reach near 25th
magnitude in the r-band.  The field depths, time-base and coordinates
of the fields are shown in Table 1.
Figure~\ref{fig:MapDECAM2018paper} shows the field locations on the
sky.  Table 1 has about 1050 square degrees of new fields, giving the
total surveyed area to date about 2130 square degrees when including
the fields from Sheppard and Trujillo (2016).

\section{Discovery and Orbit of 2015 TG387}

2015 TG387 was found near 80 au at Subaru on UT October 13, 2015 with a
magnitude of 24.0 in the r-band.  Surprisingly, like 2012 VP113,
2015 TG387 is relatively bright compared to the limiting magnitude of
most of the survey fields.  The Subaru observations generally go
deeper than 25.5 mags, making 2015 TG387 about 1.5 mags brighter than the
limiting magnitude of the disocovery survey fields it was found in.
At 80 au, the expected diameter of 2015 TG387 would be about 300 km
assuming a moderate albedo of 15 percent.

\begin{figure}
\epsscale{0.4}
\centerline{\includegraphics[angle=90,totalheight=0.4\textheight]{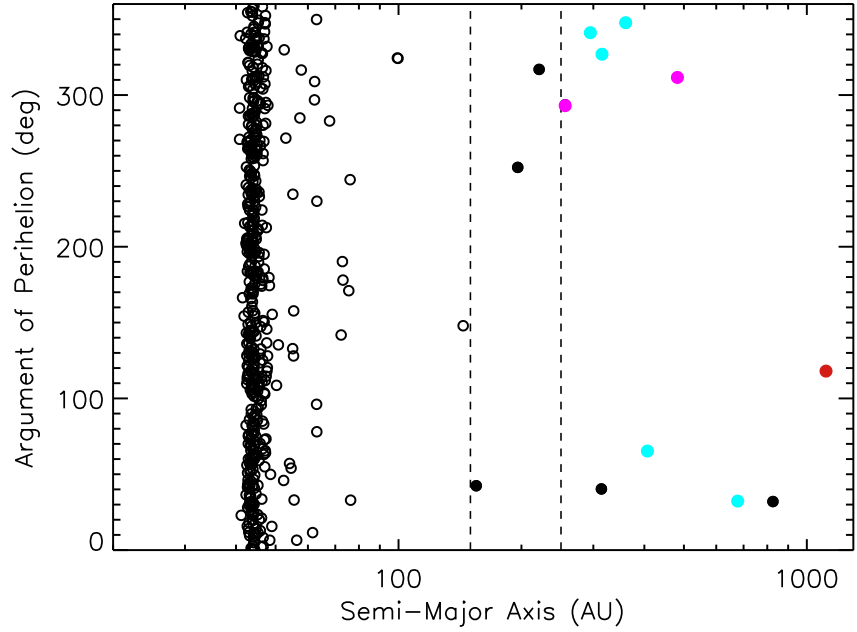}}
\caption{Semi-major axis versus argument of perihelion for all objects
  with perihelia greater than 40 au.  Colors are the same as described
  in Figure~\ref{fig:kboeq2018paper}.  There is a noticeable clustering
  between 290 and 40 degrees, of which 2015 TG387 is the first detached
  extreme or inner Oort cloud object to be closer to 180 degrees than
  0 degrees with a semi-major axis beyond 250 au.
\label{fig:kboaw2018morePaper} }
\end{figure}

2015 TG387 was recovered in December 2015; September, October,
November and December 2016; September and December 2017; and May 2018.
It thus has multiple observations during 4 separate oppositions.  The
barycentric orbital elements and their current uncertainties are shown
in Table 2.  2015 TG387 has a moderately reliable orbit.  Because the
semi-major axis has been found to be one of the largest known for a
trans-Neptunian object that always stays beyond Neptune, second only
to 2014 FE72 (Sheppard and Trujillo 2016), the uncertainties in the
orbit of 2015 TG387 are still modest even with the multiple
oppositions of observations.  We find a semi-major axis of $a=1170\pm
70$ au, eccentricity of $e=0.945\pm 0.003$, inclination of
$i=11.670\pm0.001$ deg, longitude of the ascending node of $\Omega =
300.98\pm 0.01$ deg, and argument of perihelion of $\omega = 118.2 \pm
0.1$ deg at an epoch of 2457308.8.  This makes the argument of
perihelion closer to 180 degrees than 0 degrees, which is unlike
Sedna, 2012 VP113 and most of the other ETNOs as discussed in Trujillo
and Sheppard (2014).  In fact, 2015 TG387 is the first detached ETNO
or inner Oort cloud object to have an argument of perihelion closer to
180 as opposed to 0 degrees (Figure~\ref{fig:kboaw2018morePaper}), but
as discussed below, planet X can constrain the argument of perihelion
of 2015 TG387 to be near zero degrees, like that found for the other
ETNOs and IOCs.

\subsection{Longitude of Perihelion of 2015 TG387}

2015 TG387's Longitude of Perihelion is $\bar{\omega} = \omega +
\Omega = 59.2\pm 0.2$ degrees.  Trujillo and Sheppard (2014) suggested
the inner Oort cloud objects might be asymmetric in longitude, but
believed there were significant observational biases in the longitude
discoveries of these objects and thus any longitude asymmetries would
require further unbiased extreme object discoveries.  Batygin and
Brown (2016a), using the biased longitudinal observational results of ETNOs,
suggested the longitude of perihelia of many of the ETNOs are
clustered between about 0 and 120 degrees.  Sheppard and Trujillo
(2016) further examined the longitude of perihelion similarities after
discovering more ETNOs in a low biased longitudinal survey.  Sheppard
and Trujillo (2016) concluded the asymmetry in longitude for the ETNOs
is likely real, but still marginal at about the 3 sigma level with the
additional discoveries reported in the 2016 paper and assuming all
known ETNOs have observational biases similar to their own survey.
Additional analysis by Brown (2017) using all the ETNOs with their
observational biases further finds the clustering likely real.
Shankman et al. (2017b) do not find an obvious longitude of perihelion
clustering for ETNOs, though their analysis used a limited set of
surveyed longitudes and a liberal definition of what is an ETNO.  The
Shankman et al. (2017b) results are hindered by a lack of uniform sky
coverage and low number statistics, making their results hard to
interpret.  This is the reason we are performing a very uniform and
extensive survey.

\begin{figure}
\centerline{\includegraphics[angle=90,totalheight=0.4\textheight]{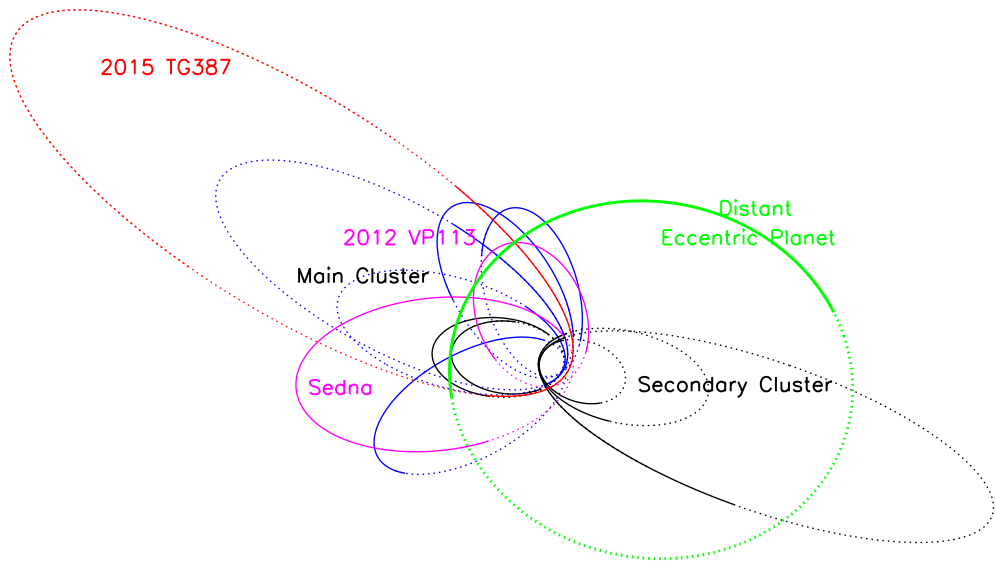}}
\caption{The plan view of the Inner Oort Cloud objects ($q>60$ au:
  Sedna and 2012 VP113 in purple), detached ETNOs (most conservative
  definition in blue with $45 < q < 50$ au and $a>250$ au: 2015 RX245,
  2014 SR349, 2013 SY99, 2010 GB174, and 2004 VN112 and less
  conservative definition in black with $40 < q < 45$ au and $a>150$
  au: 2015 KG163, 2013 UT15, 2013 GP136, 2013 FT28, 2000 CR105) and
  the newly discovered IOC 2015 TG387 in red.  All but the least
  conservative detached ETNOs are anti-aligned in longitude of
  perihelion with a possible distant planet on an eccentric orbit (in
  bold green).  There is a second grouping of ETNOs that to date
  generally have lower perihelia and appear aligned with the planet in
  longitude of perihelion.
\label{fig:ETNOplan2018} }
\end{figure}

Our survey has now covered even more sky since Sheppard and Trujillo
(2016) and our longitude biases are now even smaller.  2015 TG387 was
found about 45 degrees away in right ascension (RA) on the sky from
Sedna and 2012 VP113 and in addition, unlike Sedna and 2012 VP113, was
not found within a few au of perihelion but about 15 au away from
perihelion at around 80 au. 2015 TG387 is brighter than 25.5 magnitude
over an RA range of about 23 to 9 hours, which corresponds to
heliocentric distances less than about 100 au, after which 2015 TG387
would be too faint for us to efficiently discover. This means 2015
TG387 could easily have been discovered and/or had a longitude of
perihelion well away from the other two known inner Oort cloud
objects, but the longitude of perihelion is between Sedna and 2012
VP113.  Thus 2015 TG387 continues the longitude clustering trend seen
for the inner Oort cloud objects and ETNOs, which might be caused by a
massive planet shepherding these objects
(Figure~\ref{fig:ETNOplan2018}).

\subsection{Inner Oort Cloud Observational Simulation: Longitude of Perihelion}

Sedna, 2013 VP113 and 2015 TG387 all have similar longitudes of
perihelion (96\arcdeg, 25\arcdeg, and 59\arcdeg, respectively). These
are the only objects known with perihelion greater than 60 au, where
Neptune gravitational effects are mostly insignificant (Gomes et
al. 2008, Brasser \& Schwamb 2015).  Both 2013 VP113 and 2015 TG387 were
discovered in our survey, which attempts to have low biases in
longitude discovery by observing at all times of the year.  We can
simulate their detection statistics and biases using methods similar
to Sheppard and Trujillo (2016).  Sedna is bright enough that almost
the whole sky has been searched for such bright objects and thus Sedna
can be considered to have no discovery bias in longitude (Brown 2008,
Sheppard et al. 2011, Rabinowitz et al. 2012).  The observed longitude
of perihelion distribution given Sedna, 2015 TG387 and 2013 VP113 is
$\bar{\omega} = 60\arcdeg \pm 35\arcdeg$ which is a fairly narrow
standard deviation.

Assessing whether this observation is consistent with a uniform
distribution is not straightforward given the small number of
detections and the fact that $\bar{\omega}$ is a continuous angle
across the sky. Kuiper's one-sided variation of the Kolmogorov-Smirnov
statistic might normally be used, but its statistics are not well
defined for number of detections less than 4 (Press et al. 1992), nor
is it particularly sensitive with such a low number of detections.

To determine if the longitude of perihelion trend is statistically
significant, we first examine a simplistic case where we consider
Sedna's longitude of perihelion as an a priori value, then draw two
objects (2015 TG387 and 2013 VP113) from a uniform distribution in
$\bar{\omega}$ ignoring observational biases. If both objects were
within 35\arcdeg of Sedna's $\bar{\omega}$ then the standard deviation
of all three objects would be less than 35\arcdeg. Assuming binomial
statistics this would happen with a probability of $P' = \left(\frac{2
  \times 35\arcdeg}{360\arcdeg}\right)^2 = 0.038$, or the equivalent
of a $2.1 \sigma$ assuming Gaussian statistics. This method uses only
the 3 known IOCs and does not take observational biases into account.

\begin{figure}
\centerline{\includegraphics[angle=0,totalheight=0.4\textheight]{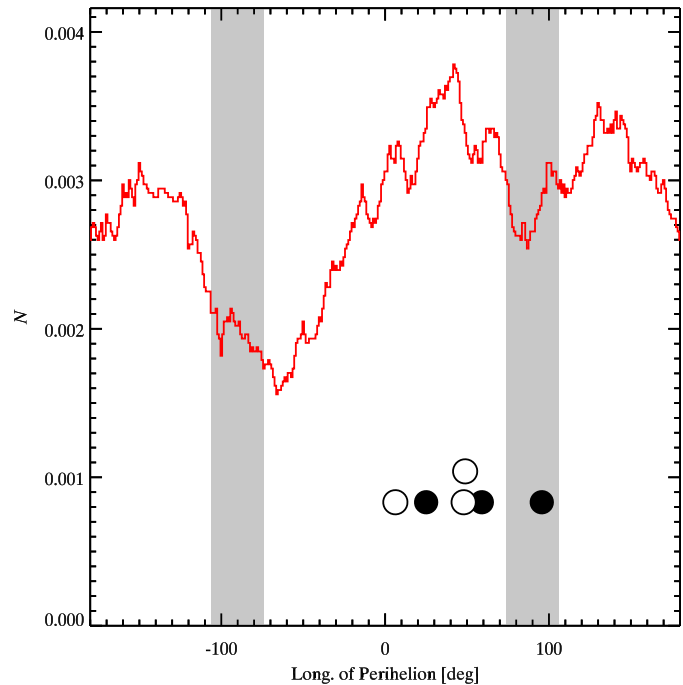}}
\caption{The normalized bias in longitude of perihelion in our survey
  fields for the IOC objects if drawn from a uniform longitude of
  perihelion distribution using the parameters shown in Table 3.  The
  ecliptic longitudes within 15\arcdeg of where the Galactic plane
  crosses the ecliptic are shaded in grey (southern declination is
  left and northern declination is right). The fact that the highest
  and lowest points differ by only about a factor 2 in all but a few
  cases suggests that our survey has fairly low observational biases
  for longitude of perihelion.  The longitude of perihelion of Sedna,
  2012 VP113 and 2015 TG387 are shown by solid circles while their
  location at discovery are shown by open circles.
\label{ioclonperiq4} }
\end{figure}

To simulate observational biases more thoroughly we assume that
2015 TG387 and 2013 VP113 are drawn from the same population of extremely
distant objects. We use the distribution detailed in Table 3 as the
underlying distribution assuming that the objects have a uniform
longitude of perihelion and argument of perihelion. We then tally the
number of objects in our observational simulation that would be
detected given our combination of survey field sky locations and
depths.  The resulting survey longitude of perihelion bias is shown in
Figure~\ref{ioclonperiq4}.  We then assess whether the distribution of
$\bar{\omega}$ for the simulated detections is consistent with the
actual observed distribution of $\bar{\omega} = 60\arcdeg \pm
35\arcdeg$.  For this we require that a randomly selected group of two
simulated detected objects together with Sedna have a standard
deviation of less than $35\arcdeg$ and that the mean also differ from
$60\arcdeg$ by less than $35\arcdeg$. We constructed a simple
Monte-Carlo simulation and found that these criteria are satisfied
about 5\% of the time by a uniform distribution, or about a $2\sigma$
significance that the IOCs are not drawn from a uniformly distributed
population in longitude of perihelion.  As seen in
Figure~\ref{ioclonperiq4} and the results above, our survey has fairly
uniform longitude of perihelion discovery statistics as including our
observational biases only slightly changes the probability, making an
asymmetric population of IOCs a little less significant.

The results in the previous paragraph are for a $q'=4$ power-law size
distribution. We also simulated a $q'=5$ power-law size distribution
and found similar results.  We conclude that although the longitude of
perihelion clustering continues to appear intriguing, with only 3
known IOCs, several more need to be discovered in low-bias longitude
surveys for this effect to be statistically significant at the
$3\sigma$ level for IOCs.

\subsection{Extreme Trans-Neptunian Object Observational Simulation: Longitude of Perihelion}

We now include the discovery statistics of the detached Extreme
Trans-Neptunian Objects (ETNOs) along with the IOCs from our
survey. The detached ETNOs can be observationally differentiated from
the IOCs because of their more moderate perihelia ($40-45 \mbox{ au} <
q < 50-60 \mbox{ au}$) compared to the IOCs' extremely high perihelia
of $q > 50-60$ au.  The detached ETNOs may have similar formation and
evolutionary histories as the IOCs, but they are significantly closer
to the giant planets, and thus could have obtained their orbits
through different processes (Brasser \& Schwamb 2015, Bannister et
al. 2017).  Assuming a $q > 40-45$ au and $a > 150-250$ au definition
for the detached ETNOs, the detached ETNOs found in our survey are
safely 2014 SR349 and if less conservative also 2013 FT28 (Table 2).
We also found several ETNOs that have lower perihelia than 40 au and
thus are not detached but scattered ETNOs with more interactions with
Neptune: 2014 FE72, 2013 UH15, 2013 FS28, and 2014 SS349.  These
scattered ETNOs were detailed in Sheppard and Trujillo (2016) and here
we only consider the detached ETNOs 2014 SR349 and 2013 FT28, and in
the most conservative case we only consider 2014 SR349, with its
perihelion above 45 au and semi-major axis above 250 au.

The statistics for the longitude of perihelion clustering improve by
including 2014 SR349 with the discovery statistics computed above for
IOCs Sedna, 2015 TG387 and 2012 VP113. For this group of objects, the
mean $\bar{\omega} = 49\arcdeg \pm 36\arcdeg$. We computed the
significance of this in the simple case by running a Monte Carlo
simulation with our three detections and Sedna drawn from a uniform
distribution in $\bar{\omega}$ and ignoring observational bias. Only
1.5\% of the time do these three objects have $\bar{\omega}$ close
enough to Sedna for the mean of all objects to be within $49\arcdeg
\pm 36\arcdeg$ and the standard deviation to be $< 36\arcdeg$, the
equivalent of a $2.4 \sigma$ event in Gaussian statistics.  Using our
observational bias simulator and the method in Section 3.2, we find
that these criteria are met only $P = 2.8\%$ of the time when the
ETNOs and IOCs are drawn from a uniform $\bar{\omega}$
distribution. This is the equivalent of $2.2\sigma$ in the Gaussian
case, which again is interesting but not statistically significant at
the $3 \sigma$ level since only four objects are used in the analysis.

Using the more liberal definition in perihelion of a detached ETNO of
$q > 40$ au and $a > 250$ au would cause 2013 FT28 to be included in
the detached ETNO group as well. The orbit of 2013 FT28 has
$\bar{\omega} = -101.860\arcdeg$, which departs from the mean of the
other IOCs and ETNOs by about 180 degrees.  Sheppard and Trujillo
(2016) proposed 2013 FT28 is the first known ETNO aligned with Planet
X in longitude of perihelion.  A distant massive planet on an
elongated orbit would likely create both an anti-aligned and aligned
longitude of perihelion population of ETNOs with respect to the
Planet's longitude of perihelion (Brown and Batygin 2016a).  That is,
the aligned population has a longitude of perihelion similar to that
of the hypothesized planet on an elongated orbit while an anti-aligned
population would differ by about $180\arcdeg$.

If we make the assumption that 2013 FT28 is a member of the opposite
ETNO population than Sedna, 2012 VP112, 2015 TG387 and 2014 SR349, its
longitude of perihelion does indeed fit the $\bar{\omega}$ pattern
already discussed as $180\arcdeg - 101.86\arcdeg = 78.14\arcdeg$,
which is well within the mean $\bar{\omega}$ for the other ETNOs and
IOCs ($49\arcdeg \pm 36\arcdeg$). If this were the case, this would
decrease the probability that $\bar{\omega}$ for the ETNO and IOCs
were drawn from a random uniform distribution. From our observational
bias simulation, this would suggest $P = 0.013$ or $2.5\sigma$ in
Gaussian statistics.  Again, not yet statistically significant as
there are only a few objects used, but cause for interest.

Above we are trying to use only objects found in low longitude bias
surveys, but very few objects have been found like this and is the
main reason we are continuing our survey.  Taking all known
conservatively defined detached ETNOs and IOCs ($q>45$ au and $a>250$
au) into account finds 8 objects (Sedna, 2012 VP113, 2015 TG387, 2004
VN112, 2010 GB174, 2013 SY99, 2014 SR349 and 2015 RX245), all with
longitude of perihelion between 15 and 120 degrees.  This has a
$0.005\%$ chance of happening and is about a $4 \sigma$ event, if
ignoring possible observational biases.

We note that the above methods include a priori assumptions
identifying the location of the longitude of perihelion clustering, as
well as the specific choices of objects and the population membership
of 2013 FT28.  We additionally measured the statistical significance
of the 8 longitude of perihelion measurements using the R Package
CircStats (Jammalamadaka \& SenSupta 2001) implementation of the Kuiper
test for circularly distributed variables (result 2.2 with probability
drawn from uniform distribution $<1\%$) as well as the Rayleigh test
for uniformity (result mean length 0.84 with probability drawn from
uniform distribution 0.16\%, equivalent of $3.2\sigma$ assuming
Gaussian statistics). These tests are a little lower in significance
than the binomial test. They have the advantage that they do not
pre-suppose a specific longitude of perihelion range but the
disadvantage that they are best suited for larger datasets.  So again,
the longitude of perihelion clustering is interesting, but further low
biased discoveries are needed to make it statistically significant.

\subsection{The Inner Oort Cloud Semi-Major Axis Distribution}

We have detected two IOCs in our survey with very different semi-major
axes, 2015 TG387 with $a = 1170$ au and 2012 VP113 with $a = 270$
au. This suggests a semi-major axis distribution that has more distant
objects than the $a^{1}$ assumed by Sheppard and Trujillo (2016), a
value that was largely assumed due to limited discovery
statistics. Since solar system volume increases with heliocentric
distance as $R^3$, although a $a^{1}$ semi-major axis distribution has
more objects with large semi-major axis than small, the real
underlying space density of objects for a $a^{1}$ distribution would
fall as $R^{-2}$. According to our observational bias simulations,
assuming $a^{1}$ one would expect to find about 23 objects within 10
au of 2012 VP113's low semi-major axis for every object within 10 au
of 2015 TG387's high semi-major axis. Since the number of true detections
is only 2, we cannot statistically rule out the possibility of a
$a^{1}$ semi-major axis distribution. However, we find that to observe
equal numbers of $a = 270$ au and $a = 1170$ au objects we would have
to draw them from an $\sim a^{2.7}$ semi-major axis distribution. Such
a semi-major axis distribution would imply a much larger number of
IOCs and a population that is highly dependent on the number of very
distant objects, which are the most difficult to observe. All other
parameters remaining the same, this $a^{2.7}$ population would have to
be a factor $\sim 2$ larger than an $a^{1}$ population. We use this
$a^{2.7}$ semi-major axis distribution as our favored semi-major axis
distribution for the remainder of this work. Interestingly, if the
true semi-major axis distribution is close to power law of exponent 3,
as these simulations suggest, this implies a fairly constant space
density of objects moving outward with distance.

\subsection{Population Number and Mass of IOCs}

From the observational simulations we find that the total number of
IOC objects is quite large, mainly because IOCs can only be detected
for a small fraction of their orbits. For 2015 TG387, assuming a
semi-major axis of 1170 au and a perihelion of 65 au, 2015 TG387 will be
fainter than our faintest HSC survey field depths, $r \sim 25.5$ mags,
for 99.5\% of its orbital period.  Using our observational bias
simulation with a variety of parameters (combinations of $q=4,5$ and
$a=2.7$), the total number of objects larger than radius 20 km based
on our 2 detections (2015 TG387 and 2012 VP113) is roughly $2 \times
10^6$ with total mass of about $10^{22}$ kg. This total mass likely
exceeds that of Classical dynamically ``cold'' TNOs and is similar to
the dynamically ``hot'' TNOs which have mass of about $M = 1.8 \times
10^{21}$ kg and $6.0 \times 10^{22}$ kg, respectively (Fraser et
al. 2014). This is a lower limit on the IOC mass given that we have
almost no constraints on distant, low eccentricity objects which are
unobservable given current technology.  Sheppard and Trujillo (2016)
showed the Cumulative Luminosity Function and size distribution of the
ETNOs and IOCs are likely similar to the Kuiper Belt as well.

\subsection{Observational Bias Simulation: Inclination Distribution}

The inclinations of our IOCs and detached ETNOs are moderate, between
$11.7\arcdeg < i < 24.0\arcdeg$. Our survey observed mostly between 5
and 25 degrees from the ecliptic, though we did approach the ecliptic
and thus were sensitive to objects with inclinations significantly
lower than that detected. We can say that the inclination distribution
of the IOCs and detached ETNOs is quite thick, and in fact in our
survey simulation we use the inclination distribution for the
scattered disk objects as found by Gulbis et al. 2010 ($\mu_1 =
19.1\arcdeg$ and $\sigma_1 = 6.9\arcdeg$) as this is similar to the
average ETNO inclination found by Sheppard and Trujillo (2016). Our
observational bias simulator suggests that the observed ETNO/IOC
inclination distribution is consistent with the inclination
distribution of the scattered disk objects.

We can rule out narrow distributions for the ETNOs/IOCs. For instance,
our observations are not consistent with the more inclined component
of the Classical KBOs ($\sigma_3=8.1\arcdeg$, Gulbis et al. 2012). The
simulated detected distribution in this case has a mean inclination of
$i = 10.5\arcdeg \pm 4.8\arcdeg$. This is inconsistent with the
detection of 2012 VP113 with an inclination of $i \sim
24\arcdeg$. From our observational bias simulation, the probability of
us discovering an object with $i > 24\arcdeg$ is 0.0032, which is
ruled out at the $3 \sigma$ level assuming Gaussian statistics. Given
this and the fact that 2013 FT28 and 2014 SR349 have inclinations $i >
17$ which each have a probability of discovery of only 0.22, we can
reject the hypothesis that the ETNOs/IOCs follow the inclination
distribution of the Classical TNO populations at the $>3\sigma$ level.

We can also say that the IOCs and detached ETNOs likely do not have a
high inclination distribution like that found for TNOs with perihelia
above 40 au and experiencing both Neptune Mean Motion Resonance (MMR)
and Kozai Resonance (KR) behavior.  These MMR-KR objects have an
average $i \sim 28\arcdeg$ (Sheppard et al. 2016) and we have found no
detached ETNOs or IOCs with inclinations this high while finding
several even more inclined MMR-KR objects in the same survey.  One
very high inclination scattered ETNO has been found, 2015 BP519 with
an inclination of 54 deg, but this object has a fairly low perihelion
near Neptune and thus may not be related to the IOCs and detached
ETNOs with much higher perihelia (Becker et al. 2018).

\subsubsection{Inner Edge of the Inner Oort Cloud}
We do not know whether the Trans-Plutonian IOCs ($q > 50-60$ au) and
the ETNOs ($q < 50-60$ au) have similar origins or could be formed
from different mechanisms.  The ETNOs have low enough perihelia that
their orbits can still be modified by interactions with the known
giant planets.  2015 TG387 has such a large orbit that outside forces
can have significant effects on its orbit (see below), and thus its
orbit can also be modified, but likely in a different manner than the
ETNOs.  To date, 2012 VP113 and Sedna are the only known objects with
high enough perihelia and low enough aphelia that their orbits are not
significantly modified by any known forces, either internal forces
(the giant planets) or external (the galactic tide and passing stars).
Trujillo and Sheppard (2014) suggested an inner edge to the IOC
population and Sheppard and Trujillo (2016) expanded on this since no
IOCs or ETNOs have been discovered with perihelia between about 50 and
75 au, though they would be significantly easier to discover than the
more distant Sedna and 2012 VP113.

Some simulations of stellar encounters (Morbidelli and Levison 2004)
suggest an inner edge to the IOCs could be created as part of the
formation process. So if the ETNOs are formed from a different process
than the IOCs, the two populations could have dynamically different
origins. It is somewhat difficult to construct a single population
that can fulfill the detection statistics of both groups using our
observational bias simulator. However, this dirth of objects is only
marginally significant due to the small number of known IOCs (Sheppard
and Trujillo 2016).

With the discovery of 2015 TG387, we have now found an object that has
a perihelion well in the 50 to 75 au range, and though it is likely
stable, it is not as stable as Sedna and 2012 VP113 and thus may not
be a good object to use in accessing the inner edge of the IOC as its
perihelion may be modified by the action of the galactic tide and
passing stars.  Nevertheless, we attempted to construct a population
that could span both the ETNOs and the IOCs with a $q'=4$ size and
$a=1$ semimajor axis distribution and found the total normalized ratio
of objects with $q \leqq 50$ au, $50 \mbox{ au} < q < 65$ au, $q \geqq
65$ au was 0.34:0.45:0.21, which departs from the observed
distribution of 6:0:2. Assuming binomial counting statistics drawn
from our bias simulator, the probability of observing no IOC/ETNO
transition objects (0.45 probability per object) in the $50 \mbox{ au}
< q < 65$ au regime while finding 8 objects outside that regime (0.55
probability per object) is 0.0084, equivalent of a $2.6 \sigma$ event
assuming Gaussian statistics. We also considered the $a=2.7$ semimajor
axis power law and found similar results. In this case, the total
normalized ratio of objects with $q \leqq 50$ au, $50 \mbox{ au} < q <
65$ au, $q \geqq 65$ au was very similar with 0.21:0.46:0.33. Assuming
binomial counting statistics the probability of observing no objects
in the gap region is then 0.0072, equivalent of $2.7\sigma$ assuming
Gaussian statistics. Thus the observation of a "gap" in the 50 au to
65 au regime is not formally statistically significant at the 3
$\sigma$ level without at least two more detections, but is suggestive
of a possible observed gap, or at least a smaller number of objects,
in the population between the $q < 50$ au ETNO population and the $q >
65$ au IOC population. In addition, we note that no other surveys have
found objects in this gap region. Although we didn't specifically
simulate other surveys, this is suggestive that a true gap, or at
least a paucity of objects does exist.

If the perihelion gap between ETNOs and IOCs is confirmed to be real
through more discoveries, one possible way the gap may be formed is
through resonances with the unobserved distant planet.  We know each
distant Neptune mean motion resonance has a particular favored
perihelion distance associated with it.  The Plutinos generally come
to perihelion around 30 au, while the 5:2 resonance is 32 au, 7:4 is
38 au, and 6:1 is 39 au.  In this sense, it is possible the ETNOs and
IOCs are in resonances with the distant planet, and these resonances
don't generally prefer perihelia in the 50 to 65 au range.  Further
modeling and simulations are needed to look at this possibility,
though the complicated dynamics and the many unknowns might make this
only possible to fully analyze once an actual massive planet if found
in the outer solar system.

\section{Simulations of 2015 TG387's Orbit Stability}
We ran several numerical simulations to determine the orbit stability
of 2015 TG387 over the age of the solar system under differing
conditions.  In all simulations, we used the nominal orbit of 2015 TG387
and clones within 3 sigma of the nominal orbit when including the
orbital uncertainties.

\subsection{2015 TG387's Stability With Known Major Planets}
\begin{figure}
\vspace*{1.2in}
\centerline{\includegraphics[angle=0,totalheight=0.45\textheight]{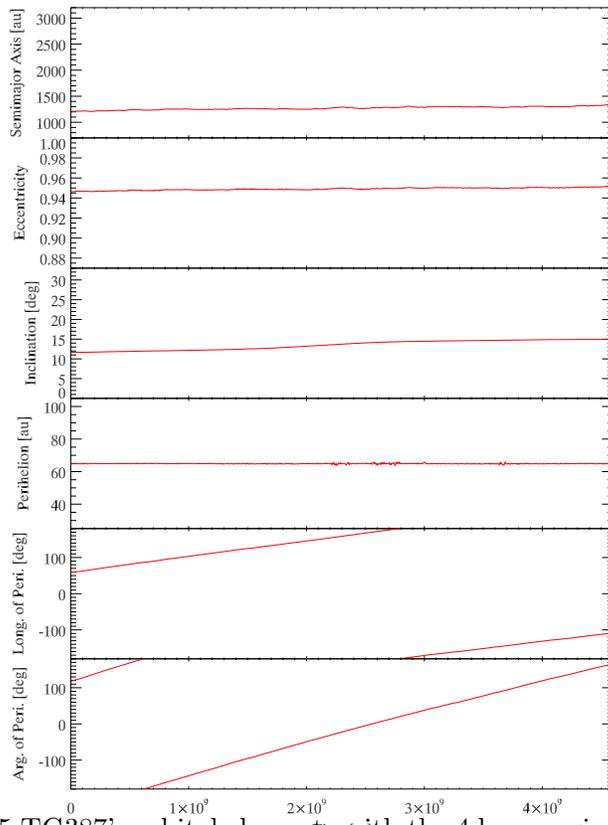}}
\caption{Evolution of 2015 TG387's orbital elements with the 4 known
  giant planets. The orbital elements for the nomial 2015 TG387 orbit is
  shown in red. Note that the longitude of perihelion and the argument
  of perihelion cycle through $\sim 360\arcdeg$ over the age of the
  solar system due to the action of the known major planets.
\label{2015TG387sim} }
\end{figure}

We found the orbit of 2015 TG387 is very stable when including only the 4
giant planets and the Sun in our numerical simulations using the
Mercury program (Chambers 1999).  2015 TG387's semi-major axis,
eccentricity and inclination change little over the age of the solar
system (Figure~\ref{2015TG387sim}).  2015 TG387's longitude of perihelion,
argument of perihelion and longitude of the ascending node cycle or
precess through 360 degrees from the minor interactions with the
quadrupole of the solar system's potential.  For 2015 TG387, the
time-scale for the argument of perihelion and longitude of perihelion
cycle is about 4 and 6.5 Gyr, respectively, when just including the 4
giant planets.  This precession is true for all ETNOs that remain
beyond Neptune for their entire orbit, though the cycle time-scales
differ depending on the object semi-major axis and perihelion distance
as an object with higher semi-major axis and/or higher perihelion will
have less interactions with the known giant planets and thus take
longer to precess (Trujillo and Sheppard 2014). See below for a
further discussion of the longitude of perihelion and the ETNOs'
precession about these angles for various different simulations.

\subsection{2015 TG387's Orbit with Galactic Tide}

To better understand the long-term dynamical behavior of 2015 TG387 to
forces outside of the Solar System, we integrated 100 clones of 2015
TG387 in several simulations with the presence of the known giant
planets and Galactic tide for 4 Gyrs.  We directly integrated the four
giant planets as active bodies.  The time step was 200 days.  The
Galactic tidal model used the same formulation as Levison et
al. (2001).  In this model, the vertical tide is $\sim 1$ order of
magnitude stronger than the radial component. This overall strength of
the tide is largely set by the local density of matter in the Galactic
disk, which we set to 0.1 solar masses per cubic parsec (Holmberg \&
Flynn 2000).

An example of the evolution of one of these clones with the Galactic
tide included is shown in Figure~\ref{KaibGT}. We also compare it with
the evolution of Sedna, 2012 VP113, and 2014 FE72. As can be seen in
the figure, the perihelion of 2015 TG387 can fluctuate by $\sim \pm
10$ au over the course of 4 Gyrs due to influence from the Galactic
tide. When the 2015 TG387 clone is driven down to a perihelion of
$\sim$60 AU, the object begins receiving small energy kicks from the
giant planets, driving a minor diffusion in its semi-major axis, which
then restabilizes when the perihelion rises above $\sim$60 AU
again. In contrast, Sedna has very small changes in perihelion as its
aphelion is only about 1000 au while 2012 VP113 undergos almost no
changes in perihelion due to its much smaller semi-major axes, which
limits the effects of the Galactic tide. Their non-evolving, larger
perihelia result in semi-major axes that are virtually fixed with
time. On the other hand, 2014 FE72 has a much smaller perihelion and a
larger semi-major axis than 2015 TG387. Within a couple hundred Myrs,
its perihelion has moved even closer to the planets leading to its
ejection from the solar system.

The behavior of 2014 FE72 demonstrates that if 2015 TG387 evolves too
much in perihelion its semi-major axis can be inflated via planetary
perturbations to the point where it becomes unstable due to wild
swings in pericenter driven by the Galactic tide when the semi-major
axis is large. However, our integrations of clones demonstrate that
such behavior is unlikely for 2015 TG387 as it has a significantly high
perihelion and low semi-major axis to prevent major changes to its
orbit from both inside and outside forces. Out of our 100 clones, only
one was ejected from the solar system over 4 Gyrs of evolution (a
clone near the $3\sigma$ uncertainty limit), and during the last Gyr
of integration any given clone only had a 0.8\% chance of having
$q<55$ AU.

When including the outside forces from the Galactic tide with the
known major planets in our simulations, 2015 TG387's orbit is mostly
stable, though there is a little more movement in its semi-major axis,
eccentricity and inclination from some minor interactions with the
Galactic tide than without the Galactic tide simulations.  Only a few
percent of the clones have significant movement of some 10 au in
perihelion and 100 au in semi-major axis over 4 Gyrs, but all but one
have their perihelia still above 55 au.  We consider 2015 TG387 to be
stable to the Galactic tide.

\subsubsection{The Combination of the Galactic Tide and Solar System's Quadrupole Moment on Orbit Stability of ETNOs}

\begin{figure}
\vspace*{0.5in}
\centerline{\includegraphics[angle=0,totalheight=0.4\textheight]{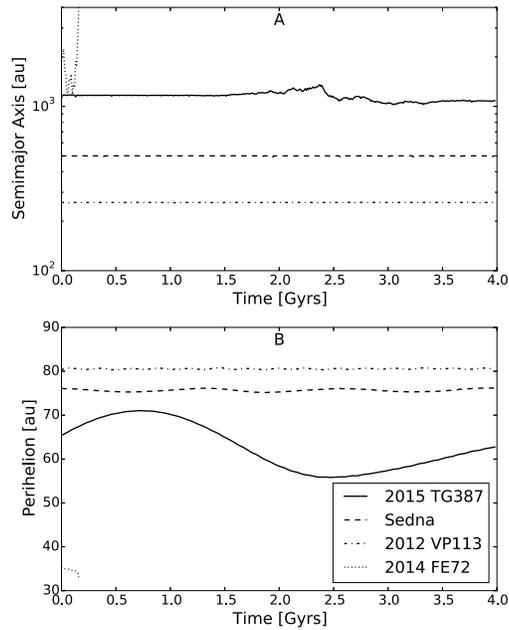}}
\caption{The behaviour of different extreme TNOs to the Galactic tide
  and 4 giant planets.  2015 TG387 is stable to the Galactic tide,
  though its perihelion now varies between about 55 and 70 au from
  Galactic tide interactions.  Sedna, with a lower semi-major axis but
  aphelion still near 1000 au, has a slight variation in perihelion
  from the Galactic tide, while 2012 VP113, even more tightly bound,
  has almost no perihelion variation.  2014 FE72 has a larger
  semi-major axis and has its perihelion quickly driven down into the
  giant planet region by the Galactic tide and is lost.  Notice how
  2015 TG387's semi-major axis only starts to fluctuate when its
  perihelion is below about 60 au.  This is from energy kicks from the
  giant planets.}
\label{KaibGT}
\end{figure}

Our above results show that outside forces work in tandem with inside
forces when an object has a perihelion less than 60 au and a
semi-major axis above 1000 au.  The amount the Galactic tide affects
an ETNOs orbit is a complicated function of the semi-major axis,
inclination and perihelion distance of the object.  The higher the
semi-major axis the more the Galactic tide becomes important, with
objects beyond 1000 au starting to have significant interactions with
the Galactic tide as shown by Sedna in Figure~\ref{KaibGT} (see also
Duncan et al. 2008; Kaib \& Quinn 2009; Soares \& Gomes 2013).  If
just the Galactic tide was important, one could calculate an object's
perihelion change over time based on the object's semi-major axis and
inclination (see Heisler \& Tremaine 1986).  But the perihelion
distance is also very important for the stability of an object as once
it drops below about 60 AU, interactions with the known giant planets
become important as shown by the simulations of 2015 TG387's orbit
(Figure~\ref{KaibGT}). These interactions deliver weak energy kicks
driving a diffusion in semimajor axis. If the semimajor axis
increases, then the Galactic tide can drive the perihelion even closer
to the planets, generating stronger energy kicks. If the perihelion
drops near or below 30-35 au, the object will likely become unstable
from strong gravitational interactions with the giant planets, like
found for 2014 FE72.

Thus an object with a higher perihelion but similar distant semi-major
axis is more stable since the object would have similar Galactic tide
interactions from outside forces but less interactions with the known
giant planets.  But there is also the precession effect from the
quadrupole moment of the giant planets interaction to consider.  An
object with faster solar system precession of its orbital angles,
most importantly its argument of perihelion as this angle determines
the sign of the Galactic tide perturbation, makes the Galactic tide
perihelion perturbation shift from positive to negative and back over
time, having a canceling effect on the Galactic tide.  The cycling of
the argument of perihelion from just the giant planets can be seen in
Figure ~\ref{2015TG387sim}, while the fluctuation of the perihelion from
the Galactic tide going positive to negative can be seen in
Figure~\ref{KaibGT}.

A slow precession of the argument of perihelion, with the critical
period being longer than the age of the solar system, will not allow
the precessing to suppress net Galactic tide perturbations and thus
the Galactic tide perturbations just become larger over time, so tidal
shifts get larger causing the perihelion to have larger movement.  An
object will have slower precession with a higher perihelion or higher
semi-major axis since it will interact less with the known inner giant
planets.

2015 TG387 is near the limits of stability, but appears to be a mostly
stable object.  Its semi-major axis and aphelion distance are large
enough to cause significant interactions with outside forces, but its
perihelion is just high enough that the fluctuations of its perihelion
keep it from strongly interacting with the giant planets.  If 2015 TG387
had a little lower perihelion distance or little higher semi-major
axis, it would be a much more unstable orbit from the combination of
outside and inside forces.

\subsection{2015 TG387's Orbit with Galactic Tide and Passing Stars}

We further ran several simulations that included nearby passing stars
as well as the galactic tide and 4 known giant planets for 4 Gyrs.
The passing star parameters were chosen to be similar as that found in
Rickman et al. (2008) and were meant to mimic the conditions of the
solar neighborhood.  The process of generating random field stars has
an inherent element of stochasticity, and the most powerful few
stellar encounters will tend to dominate the effects for any large set
of encounters (Kaib, Roskar, \& Quinn 2011).  We generated four
different sets of field star encounters and integrated our 2015 TG387
clones under each set separately.  These four simulations are just a
small subset of possible stellar encounter scenarios that could have
occurred over the age of the solar system, but gives us a basic
understanding of how stable 2015 TG387 is to such encounters.  While
the range of outcomes for stellar flybys is very large, each of our 4
Gyr simulations represent a compilation of the effects of some 64,000
encounters, as our stellar flyby simulations average about 16
encounters per Myr, which is near the rate inferred from recent Gaia
data (Bailer-Jones et al. 2018). Thus, our four simulations represent
about a quarter of a million stellar encounters.

\begin{figure}
\vspace*{0.5in}
\centerline{\includegraphics[angle=0,totalheight=0.4\textheight]{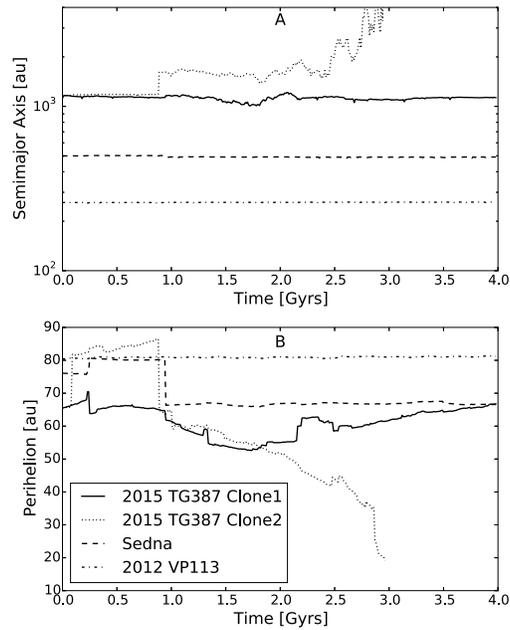}}
\caption{Same as Figure~\ref{KaibGT}, but now also including stellar
  encounters.  2015 TG387 Clone 2, Sedna and 2012 VP113 are shown
  evolving in the strongest of our four stellar encounter simulations.
  Still $35\%$ of 2015 TG387 clones survive the age of the solar
  system.  Both Sedna and 2012 VP113 also have significant changes in
  their perihelia during the strongest stellar encounter simulation.
  In our other stellar encounter simulations some $99\%$, $95\%$, and
  $94\%$ 2015 TG387 clones survive.  2015 TG387 Clone 1 has only small
  changes to its orbit over the age of the solar system in the
  moderate stellar encounter scenario that $95\%$ of the 2015 TG387
  clones survive.  We find overall 2015 TG387 is likely stable to most
  stellar encounter scenarios over the age of the solar system.}
\label{KaibGT_ST_Extreme}
\end{figure}

Unlike in the simulations with only the giant planets and the Galactic
tide, with stellar encounters we see some of the clones have more
significant evolution.  The four different simulations gave somewhat
different results as our clone survival rates after 4 Gyrs of
integration for each stellar passage simulation were 99\%, 95\%, 94\%
and 35\%.  Thus we do find there is additional pericenter evolution
from passing stars that drives more of our clones into interacting
with the giant planets, but the median case still has the vast
majority of clones surviving.  If we select the 95\% survival
simulation as our fiducial case, at any given point in the last Gyr of
integration a clone has a $\sim$16\% chance of having $q<50$
AU. However, the majority (63\%) of our clones maintained a perihelion
larger than 50 AU for the entire last Gyr of the integration,
indicating that under many stellar passage scenarios, the orbit of
2015 TG387 can remain stable for the age of the solar system.  2015
TG387 Clone 1 in Figure~\ref{KaibGT_ST_Extreme} displays the evolution
with the moderate stellar encounter scenario where 95\% of the 2015
TG387 clones survived, showing minimal changes in the orbit of 2015
TG387 Clone 1 over the age of the solar system.

The other object's evolution shown in Figure~\ref{KaibGT_ST_Extreme}
come from the simulation with the most powerful set of stellar
encounters (35\% of 2015 TG387 clones survived).  For 2015 TG387 Clone 2,
powerful stellar encounters alter the orbit around $\sim$100 Myrs and
$\sim$900 Myrs (Figure~\ref{KaibGT_ST_Extreme}).  After this, a
combination of perturbations from additional stellar passages and the
Galactic tide drive Clone 2's perihelion into the planetary region
where it begins receiving strong energy kicks from the giant
planets. This ultimately leads to its ejection after $\sim$3 Gyrs. In
this particular simulation, such behavior like Clone 2's is not rare,
as only 35\% of our clones survive the simulation. However, as
previously noted, each set of stellar encounters is unique and their
overall cumulative effect is strongly dependent on the few most
powerful stellar passages. One can see this when we study the
evolution of Sedna and 2012 VP113 in the same simulation
(Figure~\ref{KaibGT_ST_Extreme}). Although Sedna and 2012 VP113 have
more strongly bound orbits to the Sun and are generally assumed to be
very stable, a particularly powerful set of encounters can elicit some
evolution (of order $\sim$5 AU) in their perihelia, as was found in
Kaib et al. (2011).

\subsection{2015 TG387's Orbit with a distant Planet X}

Next we ran simulations to determine if 2015 TG387 could be stable to the
possible distant unknown massive planet beyond a few hundred au that
may be shepherding the ETNOs into similar types of orbits (Trujillo
and Sheppard 2014, Batygin and Brown 2016a).  In order to identify
where the planet might be in the sky, Trujillo (2019) ran thousands of
simulations of a possible distant planet using the orbital constraints
put on this planet by Batygin and Brown (2016a).  The simulations
varied the orbital parameters of the planet to identify orbits where
known ETNOs were most stable.  Trujillo (2019) found several planet
orbits that would keep most of the ETNOs stable for the age of the
solar system.

\begin{figure}
\vspace*{1.2in}
\centerline{\includegraphics[angle=0,totalheight=0.45\textheight]{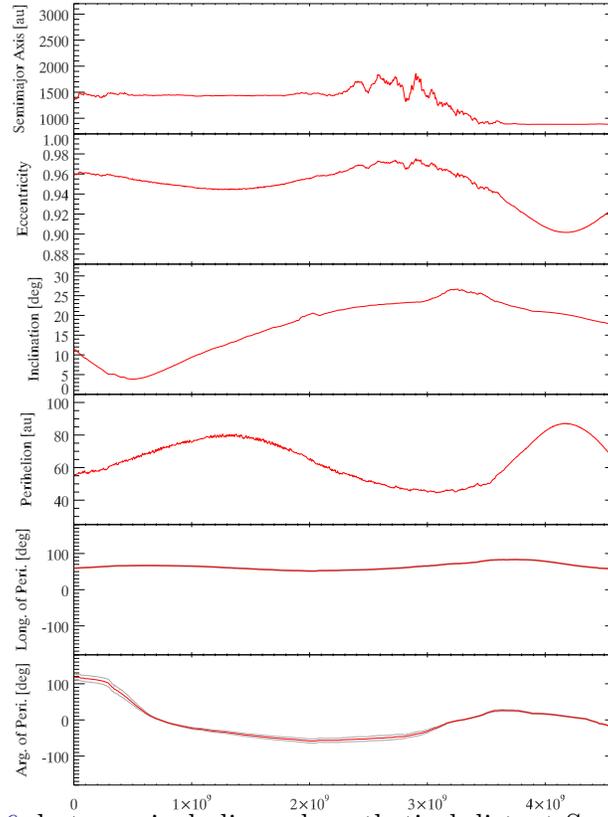}}
\caption{Same as Figure~\ref{2015TG387sim}, but now including a
  hypothetical distant Super-Earth planet on an eccentric and inclined
  orbit beyond a few hundred Astronomical Units. A typical 2015 TG387
  orbit is shown in red. In this simulation, the distant planet
  confines both the longitude of perihelion and the argument of
  perihelion of 2015 TG387, unlike without the planet in
  Figure~\ref{2015TG387sim}.  In particular, the longitude of perihelion
  of 2015 TG387 stays about 180 degrees away from, or anti-aligned with
  the planet for the age of the solar system.  2015 TG387's argument of
  perihelion also generally stays near 0 degrees after initially being
  much higher.
\label{2015TG387planetx} }
\end{figure}

To see if 2015 TG387 would also be stable to a distant planet when the
other ETNOs are stable, we used several of the best planet parameters
found by Trujillo (2019).  In most simulations involving a distant
planet, we found 2015 TG387 is stable for the age of the solar system
when the other ETNOs are stable.  This is further evidence the planet
exists, as 2015 TG387 was not used in the original Trujillo (2019)
analysis, but appears to behave similarly as the other ETNOs to a
possible very distant massive planet on an eccentric orbit.  In
Figure~\ref{2015TG387planetx} we depict the evolution of 2015 TG387's
orbit with the most favorable distant giant planet orbit from Trujillo
(2019), which is simulation number 702b/p335 in Trujillo (2019).  The
planet X parameters used in the simulation shown in
Figure~\ref{2015TG387planetx} were $a=721$ au, $e=0.55$, $i=28.75$
deg, $\omega = 142$ deg, $\Omega = 93$ deg, $M=163$ deg and a mass of
10 Earth masses.  The simulation had a timestep of 8 days and also
included the Sun and the 4 known giant planets.

\subsubsection{Longitude of Perihelion Coupled to Planet X}
The longitude of perihelion angle for a distant object beyond Neptune
will generally precess from the minor gravitational interaction of the
object with the quadrupoles of the known major planets.  The time to
make one complete revolution through the longitude of perihelion
angles depends on the object perihelion distance and semi-major axis.
The more distant an object's perihelion and/or semi-major axis the
slower the object will precess in longitude of perihelion.  It takes
about 1.3 Gyr for 2012 VP113's longitude of perihelion to precess 360
degrees, while it takes 3.0 Gyr for Sedna.  We find 2015 TG387 takes
about 6 billion years to precess 360 degrees in longitude of
perihelion when taking just the known major planets into account.  The
hypothetical distant massive planet with a perihelion of $\sim 200$ au
and semi-major axis of $\sim 700$ au precesses very slowly in
longitude of perihelion because of its distant perihelion and
semi-major axis, moving only about 15 degrees over the age of the
solar system in longitude of perihelion.

In our simulations where 2015 TG387 was stable with the additional
massive distant unknown planet, we found that 2015 TG387 frequently does
not precess on a 6 billion year time-scale in longitude of perihelion
but librates near its current longitude of perihelion, keeping it
mostly anti-aligned with the hypothetical planet for the age of the
solar system.  This keeps the object away from crossing the planet's
orbit and thus stable.  Amazingly, these distant planet orbits were
not chosen to make 2015 TG387 stable, but were simply the planet orbits
found in Trujillo (2019) that kept the other ETNOs mostly stable over
the age of the solar system.  2015 TG387 was only added in the
simulations after its orbit was well determined, which was after the
Trujillo (2019) simulation results were determined.

This is surprising to find that the third known inner Oort cloud
object is stable with the distant massive planet orbits found for the
other inner Oort cloud objects and ETNOs, and not only is 2015 TG387
stable, but it has resonance behavior with its longitude of
perihelion, librating 180 degrees from the planet's.  Planet X also
constrains the argument of perihelion of 2015 TG387.  As seen in
Figure~\ref{2015TG387planetx}, Planet X keeps the argument of
perihelion of 2015 TG387 mostly near 0 degs, just like the clustering
seen for the other IOCs and ETNOs first noticed by Trujillo and
Sheppard (2014).  Though this does not prove the hypothetical distant
planet first realized by Trujillo and Sheppard (2014) and further
revealed by Batygin and Brown (2016a) is real, it is strongly
suggestive.

\begin{figure}
\vspace*{1.2in}
\centerline{\includegraphics[angle=0,totalheight=0.45\textheight]{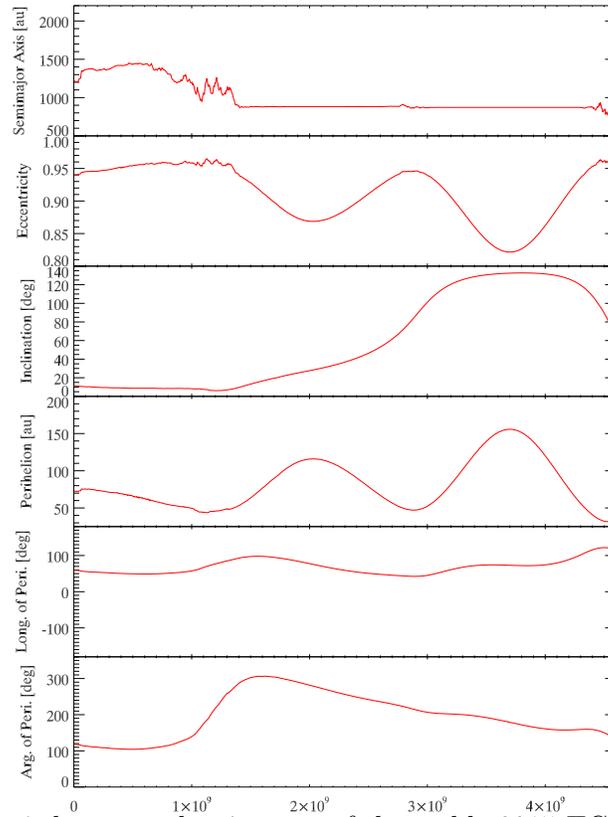}}
\caption{Same as Figure~\ref{2015TG387planetx}, but now showing one of
  the stable 2015 TG387 clones that went retrograde during the
  simulation.  Retrograde 2015 TG387 clones occurred in all of the
  simulations that had stable ETNOs for the age of the solar system
  with a planet X.  The longitude of perihelion is still constrained
  to remain anti-aligned with the planet for the age of the solar
  system.
\label{2015TG387retro} }
\end{figure}

\begin{figure}
\vspace*{1.2in}
\centerline{\includegraphics[angle=0,totalheight=0.45\textheight]{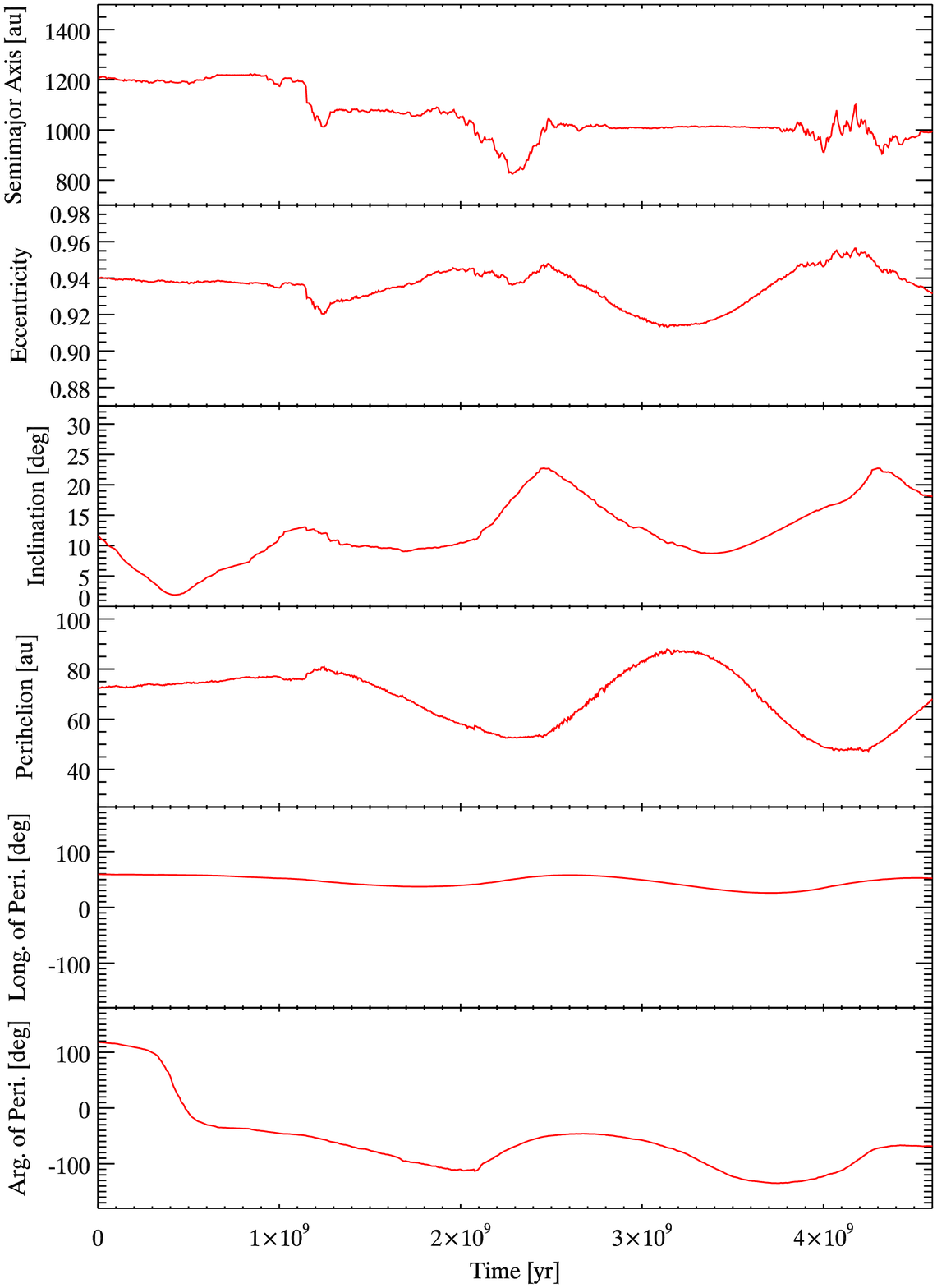}}
\caption{ Same as Figure~\ref{2015TG387planetx}, but using a retrograde
  distant planet X orbit instead of a prograde orbit. All other
  aspects remain identical including clone choice and Planet X orbital
  elements not related to the change of Planet X from prograde to
  retrograde. As with the prograde case, the distant retrograde Planet
  X confines both the longitude of perihelion and the argument of
  perihelion of 2015 TG387. Other aspects of 2015 TG387's stability remain
  similar as well in both the prograde and retrograde Planet X
  scenarios.
\label{2015TG387retro2} }
\end{figure}

We also found that in most of the dynamical simulations with planet X,
some 2015 TG387 clones became retrograde while still being in a stable
orbit in resonance with the hypothesized planet with the longitude of
perihelion still constrained to be anti-aligned with the planet
(Figure~\ref{2015TG387retro}).  Thus finding retrograde ETNOs could be a
further signature of planet X if they too are clustered into certain
orbital configurations.  This also suggests the distant planet itself
could be on a retrograde orbit.  To test this theory, we reran all of
our simulations that involved Planet X, but this time put the distant
planet on a retrograde orbit, with all other variables being the same.
This involved changing the Planet X orbital elements inclination to
$i_{retrograde} = 180^{\circ} - i_{prograde}$, longitude of the
ascending Node to $\Omega _{retrograde} = 180^{\circ} -
\Omega_{prograde}$, argument of perihelion to $\omega _{retrograde} =
180^{\circ} - \omega_{prograde}$, and mean anomaly to $M_{retrograde} =
-M_{prograde}$.  We found that 2015 TG387 behaves very similar to a
retrograde Planet X as it does to a prograde Planet X
(Figure~\ref{2015TG387retro2}).  2015 TG387 continues to be stable and
confined in longitude of perihelion and argument of perihelion angles
even with a retrograde Planet X.  We found this true for all the ETNOs.

\subsection{Stability of 2015 TG387 with Planet X, Galactic Tides and Passing Stars}

When adding in Galactic tides with the simulations involving 2015 TG387
and a distant planet, we find most clones of 2015 TG387 continue to
librate in longitude of perihelion.  Thus Galactic tides are not a
destabilizing force to 2015 TG387's libration in longitude of perihelion
with the planet.

In the passing stars case, where outside perturbations can be much
stronger and stochastic, we also found a large percentage of 2015 TG387
clones librating in longitude of perihelion with the planet for the
age of the solar system.  In a 4-Gyr simulation using our fiducial
$95\%$ 2015 TG387 clone survival stellar encounter set with the Galactic
tide and a distant planet, we find that the survival fraction of
2015 TG387 clones drops to 72\%. However, of those surviving clones, 68\%
have a longitude of perihelion that is $\pm 45^{\circ}$ of being
exactly anti-aligned with the planet's longitude of perihelion $\sim
180^{\circ}$ away, showing the surviving 2015 TG387 clones are the ones
that have longitude of perihelion resonance with the planet as found
in the previous Planet X simulations.

\section{Summary}

We discovered a new Trans-Plutonian or Inner Oort Cloud object (IOC),
2015 TG387, that has the third highest perihelion of any known object
to date at $65\pm 1$ au.  2015 TG387 was discovered in our ongoing
survey for objects beyond 50 au, which has now covered 2130 square
degrees of sky mostly using the Subaru 8m and Blanco 4m telescopes.
Assuming a moderate albedo, the diameter of 2015 TG387 is about 300
km.  The details on this new discovery are:

1) 2015 TG387's longitude of perihelion of $59$ degrees is similar to
Sedna, 2012 VP113 and the other ETNOs.  Using only the low
observationally biased discovered IOCs and detached ETNOs from our
survey as well as Sedna finds the longitude of perihelion clustering
only about a 2 to 2.5 $\sigma$ significance.  The significance of the
clustering is not at the $3 \sigma$ level because only four objects
are being used that have low observational biases in their longitude
discovery (Sedna, 2012 VP113, 2015 TG387 and 2014 SR349).  Using all 8
of the known IOCs and detached ETNOs gives a significance of $\sim
3-4\sigma$ in longitude of perihelion clustering, but this ignores the
longitude biases in the discovery of many of these objects.  Several
more IOC and detached ETNOs need to be discovered in uniform longitude
surveys to obtain a good statistical analysis of the population's
longitude of perihelion clustering.  The longitude of perihelion
clustering continues to be an interesting trend to watch.

2) With the discovery of 2015 TG387, we find the semi-major axis
distribution of the IOCs is likely an $a^{2.7}$ distribution. That is,
there are many more IOCs with high semi-major axes than low semi-major
axes.  If the power law slope of the semi-major axis distribution is
near 3 as we suggest, this implies a fairly constant space density of
objects moving outwards with distance since the volume of space goes
as the cube of distance.

3) The total number of the IOCs larger than 40 km in diameter is about
$2\times 10^{6}$, giving a total mass of about $10^{22}$ kg.  This
makes the IOC population similar in mass to the Kuiper Belt population
when using a size distribution as shown in Sheppard and Trujillo
(2016).

4) The IOCs and detached ETNOs appear to have an inclination
distribution similar to the scattered disk population of TNOs, with an
average inclination around 19 degrees.  The IOCs do not appear to have
a narrow (like the classical KBOs) or very thick (like the MMR-KR
TNOs) inclination distribution as most have inclinations between about
10 and 25 degrees inclination.

5) 2015 TG387 has a very stable orbit when just simulating the known
planets in our solar system.  The orbit is also fairly stable when
including the galactic tide.  Including passing stars over the age of
the solar system finds 2015 TG387 usually stable as well, but it is
dependent on the stellar encounter scenario used.  In most stellar
encounter scenarios, some $95 \%$ of 2015 TG387 clones are stable for the
age of the solar system.  But in the strongest stellar encounter
scenario used, some $65 \%$ of 2015 TG387 clones are lost over the age of
the solar system.  Overall, 2015 TG387 appears to be on an orbit that
most likely has lasted the age of the solar system near its current
orbital parameters.

6) We find the outside force of the Galactic tide and inside force of
the quadrupole moment of the solar system work in tandem to effect
2015 TG387's orbit.  The Galactic tide becomes important beyond about
1000 au, as seen in Sedna's orbit evolution.  The quadrupole moment of
the solar system is important when the precession of the angle of
argument of perihelion of an object's orbit is slow enough that the
Galactic tide perturbations continually increase until an object's
perihelion is pushed interior to about 60 au.  Once an object has a
perihelion interior to $\sim 60$ au, significant energy kicks from the
known giant planets cause an object's semi-major axis to change.  This
change in semi-major axis can have increased energy kicks from the
giant planets that can further lower the perihelion of an object until
it starts to more strongly interact with the giant planets.  Once the
perihelion of an object is around 30-35 au or lower, the object will
likely become unstable from gravitational scattering off the giant
planets.

2015 TG387 is near the edge of stability because it is near the
perturbations from both inside and outside forces.  That is, 2015 TG387's
perihelion is just high enough that interactions with the giant
planets are not significant, though its semi-major axis is just large
enough that Galactic tide perturbations can push 2015 TG387's perihelion
a little interior to 60 au causing slight semi-major axis variations.
But as 2015 TG387's argument of perihelion precesses from interactions
with the quadrupole moment of the solar system, the perihelion of
2015 TG387 will rise back above 60 au before any significant changes in
its orbit from giant planet energy kicks while it has a perihelion
below 60 au.  If 2015 TG387 had a little closer perihelion or little
larger semi-major axis, it would be more unstable as it would be
pushed into interacting more with the giant planets.  If an object
gets its perihelion pushed down to around 30-35 au or lower, it will
become unstable from strong gravitational interactions with the giant
planets. Thus objects that have moderate semi-major axes and are
generally thought to be stable in the inner Oort cloud (1000 to 2000
au) can still become unstabilized fairly easily if they have perihelia
that can be pushed well below about 60 au from a combination of
outside and inside forces.  The main cause of an ETNO or IOC obtaining
an unstable orbit is the strong interactions it can have with the
giant planets when its perihelion is eventually pushed to around 30 au
or less.

7) When including a massive Planet X at several hundred astronomical
units as predicted by Trujillo and Sheppard (2014) with the eccentric
and inclined rudimentary orbit proposed for it by Batygin and Brown
(2016a) in our simulations, we find 2015 TG387 is usually stable to
such a planet when the other IOCs and ETNOs are also stable.
Amazingly, in most simulations with a Planet X, we found 2015 TG387
librates in its longitude or perihelion, keeping it anti-aligned and
thus stable with the eccentric Planet X for the age of the solar
system.  This longitude of perihelion libration is not seen in the
simulations without a Planet X.  Further, Planet X also constrains the
argument of perihelion of 2015 TG387, and actually keeps it mostly
near 0 degs in our simulation shown in Figure 9, just like the current
argument of perihelia of the other known IOCs and ETNOs. These results
support the theory that a Planet X exists as 2015 TG387's orbit was
only determined after the basics of the Planet X orbit was realized,
yet 2015 TG387 reacts with the planet very similarly to the other
known IOCs and ETNOs.  In addition, some 2015 TG387 clones obtain
retrograde orbits yet still remain stable and anti-aligned with planet
X for the age of the solar system, suggesting retrograde ETNOs should
exist in most planet X scenarios.  We further found that the planet
itself might be on a retrograde orbit as 2015 TG387 and other ETNOs
were similarly stable as in the prograde planet case.

\section*{Acknowledgments}

Based in part on data collected at Subaru Telescope, which is operated
by the National Astronomical Observatory of Japan.  Observations were
partly obtained at Cerro Tololo Inter-American Observatory, National
Optical Astronomy Observatory, which are operated by the Association
of Universities for Research in Astronomy, under contract with the
National Science Foundation.  This project used data obtained with the
Dark Energy Camera (DECam), which was constructed by the Dark Energy
Survey (DES) collaborating institutions: Argonne National Lab,
University of California Santa Cruz, University of Cambridge, Centro
de Investigaciones Energeticas, Medioambientales y
Tecnologicas-Madrid, University of Chicago, University College London,
DES-Brazil consortium, University of Edinburgh, ETH-Zurich, University
of Illinois at Urbana-Champaign, Institut de Ciencies de l'Espai,
Institut de Fisica d'Altes Energies, Lawrence Berkeley National Lab,
Ludwig-Maximilians Universitat, University of Michigan, National
Optical Astronomy Observatory, University of Nottingham, Ohio State
University, University of Pennsylvania, University of Portsmouth, SLAC
National Lab, Stanford University, University of Sussex, and Texas
A\&M University. Funding for DES, including DECam, has been provided
by the U.S. Department of Energy, National Science Foundation,
Ministry of Education and Science (Spain), Science and Technology
Facilities Council (UK), Higher Education Funding Council (England),
National Center for Supercomputing Applications, Kavli Institute for
Cosmological Physics, Financiadora de Estudos e Projetos, Fundação
Carlos Chagas Filho de Amparo a Pesquisa, Conselho Nacional de
Desenvolvimento Científico e Tecnológico and the Ministério da Ciência
e Tecnologia (Brazil), the German Research Foundation-sponsored
cluster of excellence "Origin and Structure of the Universe" and the
DES collaborating institutions.  NAK received funding from NASA's
Emerging Worlds program (Grant 80NSSC18K0600) and computational
support through the OU Supercomputing Center for Education and Re-
search (OSCER) at the University of Oklahoma (OU).  This paper
includes data gathered with the 6.5 meter Magellan Telescopes located
at Las Campanas Observatory, Chile.  This research was funded by NASA
Planetary Astronomy grant NN15AF446.

\newpage

%\input{SheppardTable1b.tex}
%\documentstyle [aj_pt4]{article}    % Specifies the document style.

%\begin{document}

%\begin{center}
\startlongtable
\begin{deluxetable}{lccccc}
\tablenum{1}
\tablewidth{6.5 in}
\tablecaption{Further Inner Oort Cloud Survey Observations}
%\tablecolumns{6}
\tablehead{
\colhead{UT Date} & \colhead{Telescope} & \colhead{T}   &  \colhead{$\theta$}  &   \colhead{Limit} & \colhead{Area} \\ \colhead{yyyy/mm/dd} & \colhead{} &  \colhead{(hrs)}  &  \colhead{(``)}  &  \colhead{(m$_{r}$)} & \colhead{(deg$^{2}$)}}
\startdata
2015/10/12 & Subaru & $3-5$ & $0.45-0.7$ & 25.9  &  49.5   \\
\multicolumn{6}{c}{hsc216 23:26:44 +11:39:00} \\
\multicolumn{6}{c}{hsc204 23:30:56 +13:51:00} \\
\multicolumn{6}{c}{hsc199 23:32:20 +10:39:00} \\
\multicolumn{6}{c}{hsc292 00:04:56 +09:57:00} \\
\multicolumn{6}{c}{hsc281 00:01:20 +12:57:00} \\
\multicolumn{6}{c}{hsc302 00:07:20 +12:03:00} \\
\multicolumn{6}{c}{hsc336 00:18:20 +12:21:00} \\
\multicolumn{6}{c}{hsc367 00:26:44 +14:00:00} \\
\multicolumn{6}{c}{hsc383 00:30:44 +12:36:00} \\
\multicolumn{6}{c}{hsc410 00:38:32 +14:12:00} \\
\multicolumn{6}{c}{hsc432 00:43:56 +13:06:00} \\
\multicolumn{6}{c}{hsc448 00:48:20 +14:27:00} \\
\multicolumn{6}{c}{hsc475 00:56:20 +14:45:00} \\
\multicolumn{6}{c}{hsc481 00:58:20 +16:33:00} \\
\multicolumn{6}{c}{hsc507 01:05:20 +17:03:00} \\
\multicolumn{6}{c}{hsc527 01:12:20 +17:27:00} \\
\multicolumn{6}{c}{hsc550 01:18:32 +16:36:00} \\
\multicolumn{6}{c}{hsc568 01:23:56 +17:48:00} \\
\multicolumn{6}{c}{hsc585 01:28:44 +16:30:00} \\
\multicolumn{6}{c}{hsc625 01:39:32 +18:42:00} \\
\multicolumn{6}{c}{hsc651 01:46:44 +19:03:00} \\
\multicolumn{6}{c}{hsc689 01:57:32 +19:54:00} \\
\multicolumn{6}{c}{hsc712 02:04:08 +20:42:00} \\
\multicolumn{6}{c}{hsc727 02:08:20 +18:30:00} \\
\multicolumn{6}{c}{hsc746 02:15:20 +19:00:00} \\
\multicolumn{6}{c}{hsc755 02:17:32 +21:27:00} \\
\multicolumn{6}{c}{hsc774 02:22:32 +19:30:00} \\
\multicolumn{6}{c}{hsc786 02:26:08 +21:03:00} \\
\multicolumn{6}{c}{hsc804 02:32:08 +18:48:00} \\
\multicolumn{6}{c}{hsc808 02:33:20 +20:48:00} \\
\multicolumn{6}{c}{hsc822 02:38:32 +15:33:00} \\
\multicolumn{6}{c}{hsc825 02:39:08 +19:12:00} \\
\multicolumn{6}{c}{hsc835 02:42:08 +21:09:00} \\
\multicolumn{6}{c}{.} \\
\multicolumn{6}{c}{.} \\
\multicolumn{6}{c}{.} \\
\multicolumn{6}{c}{.} \\
\multicolumn{6}{c}{.} \\
\enddata
\tablenotetext{}{This is an abridged version, the full table and all the fields observed can be obtained at the Astronomical Journal or emailing the authors. The instruments used were HyperSuprimeCam on Subaru, DECam on CTIO4m, LB Camera on Large Binocular Telescope and IMACS on Magellan.  These are fields in our survey in addition to the fields presented and detailed in Sheppard and Trujillo (2016).  T is the approximate amount of time between the first and last images of a field, $\theta$ is the range of seeing for the night and Limit is the limiting magnitude in the r-band where we would have found at least 50\% of the slow moving objects in most of the fields.  Under the basic survey information for each night are the fields observed in J2000 coordinates for Right Ascension (hh:mm:ss) and Declination (dd:mm:ss).  Field names were the names used at the telescope for each field and are likely to be unimportant, but are included for full information.}
\end{deluxetable}
%\end{center}

%\end{document}             % End of document.

%Need to decide about my June data in galactic plane.  Yes to some of it for sure.
%Subaru June 2010
%Subaru June 2009
%Subaru June 2008
%Magell June 2008

%Schwamb another big chunk with Subaru
%43 square degrees to 25.5 to 1200 AU reference 2009, DPS, #41, 62.06   or (Schwamb, M. and Brown, M. 2009, DPS, 41, 1124)

%Also now have Kavelaars CFHT search (Chen, Y., Kavelaars, J., Gwyn,
%S., Parker, A., Suc, V., Jordan, A., Ip, W.  2012, ACM, 6244) searched
%about 100 square degrees to 25.1 in g-band with CFHTLP data.  Found
%one object considered Sedna like, though its perihelion is not quite
%Sedna like as it comes to near 50 AU.

%Fuentes another big chunck with Magellan

%CFHT Legacy survey

%CFHT OSS survey

\newpage

%\documentstyle [aj_pt4]{article}    % Specifies the document style.

%\begin{document}

\begin{center}
\begin{deluxetable}{lccccccccccc}
%\small
\tablenum{2}
\tablewidth{6.5 in}
\tablecaption{Inner Oort Cloud and Extreme Trans-Neptunian Objects \label{orbitalelements}}
\tablecolumns{12}
\tablehead{
\colhead{Name} & \colhead{$q$}  &  \colhead{$a$} & \colhead{$e$}  & \colhead{$i$} & \colhead{$\Omega$} & \colhead{$\omega$} & \colhead{$\bar{\omega}$} & \colhead{$b$} & \colhead{Dist}   & \colhead{Dia}  & \colhead{$m_{r}$} \\ \colhead{} & \colhead{(AU)} & \colhead{(AU)}  & \colhead{} &\colhead{(deg)} &\colhead{(deg)} &\colhead{(deg)} & \colhead{(deg)} & \colhead{(deg)} & \colhead{(AU)}  & \colhead{(km)}  & \colhead{(mag)} }  
\startdata
%Survey IOCs: \\
\multicolumn{12}{l}{\textbf{This Survey Inner Oort Cloud Discoveries}} \\
2015 TG387 &   $65\pm 1$  &   $1170\pm 70$    &    0.945  &    11.670  &     300.98    &   118.2  &  59.2 & 10.7   &  80.0   &  300    &   24.0  \\
2012 VP113 & 80.569   & 270.495 & 0.702 & 24.016 &  90.886 & 294.138 & 25.024 & -16.637 & 82.873 &  450 & 23.22 \\
\hline
%Other IOC: \\
\multicolumn{12}{l}{\textbf{Other Unbiased Inner Oort Cloud Objects}} \\
Sedna      &   76.167 & 540.623 & 0.859 & 11.928 & 144.502 & 311.106 & 95.608 & -11.855 & 89.593 & 1000 & 21.04 \\  
\hline
%Survey Detached ETNOs Conservative Defintion ($q>45$ au): \\
\multicolumn{12}{l}{\textbf{This Survey Detached ETNOs Conservative Defintion ($q>45$ au):}} \\
2014 SR349 & 47.483 &  290.845 & 0.836 &  17.974 &  34.839 & 341.533 &   16.372 & -17.1 & 57.2 & 200 & 24.1 \\
\hline
%Survey Detached ETNOs Liberal Defintion ($q>40$ au): \\
\multicolumn{12}{l}{\textbf{This Survey Detached ETNOs Liberal Defintion ($q>40$ au):}} \\
2013  FT28 & 43.586 &  295.912 & 0.852 &  17.381 & 217.710 &  40.430 & -101.860 &  -7.0 & 58.9 & 200 & 24.2 \\
\enddata
\tablenotetext{}{Above are the orbits of the Inner Oort Cloud Objects (IOCs) and Extreme Trans-Neptunian Objects (ETNOs) discovered in our survey. Quantities are the perihelion ($q$), semi-major axis ($a$), eccentricity ($e$), inclination ($i$), longitude of the ascending node ($\Omega$), argument of perihelion ($\omega$), longitude of perihelion ($\bar{\omega}$), ecliptic latitude at discovery ($b$), and distance at discovery (Dist). Diameter (Dia) estimates assume a moderate albedo of 0.15.  Uncertainties for 2015 TG387, if not shown explicitly, are shown by the number of significant digits. Elements for Sedna and 2012 VP113 are taken from JPL Horizons and are truncated to 3 significant digits.}
\end{deluxetable}
\end{center}

% magnitude, size

%\end{document}             % End of document.

\newpage

\begin{deluxetable}{lll}
\tablenum{3}
\tablewidth{6.5 in}
\tablecaption{Observational Bias Simulation Parameters \label{obssimparameters}}
\tablehead{
\colhead{Parameter} & \colhead{Value} & \colhead{Description}}
\startdata
\hline
& & IOC Simulation Parameters \\
\hline
$R_{\rm min}$ & 50 au            & Minimum heliocentric distance detectable in survey \\
$R_{\rm max}$ & 500 au           & Maximum heliocentric distance detectable in survey \\
$\rho$        & 1000 kg m$^{-3}$ & Density \\
$p_r$         & 0.15             & Albedo in $r$ filter \\
$r_{\rm min}$ & 20 km            & Minimum radius \\
$r_{\rm max}$ & 1200 km          & Maximum radius \\
$e_{\rm min}$ & 0.65             & Minimum eccentricity \\
$q'$          & 4                & Size distribution power law exponent \\
$\sigma_{i}$  & 6.9\arcdeg       & Sigma for Gaussian inclination distribution \\
$\mu_{i}$     & 19.1\arcdeg      & Mean for Gaussian inclination distribution \\
$a$                       & 2.7              & Semi-major axis distribution power law exponent \\
$q_{\rm min}$             & 50 au            & Minimum perihelion \\
$q_{\rm max}$             & 500 au           & Maximum perihelion distance \\
$N_{\rm obs}$             & 2                & Number of observed objects: 2012 VP113 and 2015 TG387 \\
\hline
& & Conservative ETNO + IOC simulation: same as IOC simulation except \\
\hline
$q_{\rm min}$             & 45  au           & Minimum perihelion \\
$N_{\rm obs}$             & 3                & Number of observed objects: 2012 VP113, 2015 TG387, 2014 SR349 \\
\hline
\enddata
\tablenotetext{}{Observational bias simulation parameters for both the
  IOC and ETNO+IOC combined simulation. We also ran simulations with
  both $q=5$ and $a=1$, as described in the text.}
\end{deluxetable}

\newpage

\newpage

\newpage

\newpage

\newpage

\newpage

\newpage

\newpage

\newpage

\newpage

\newpage

\begin{references}

\reference{Bai18b} Bailer-Jones, C., Rybizki, J., Andrae, R. \& Fouesneau, M. 2018, A\&A, 616, 37.
  
\reference{Bai18} Bailey, E., Brown, M., Batygin, C. 2018, AJ, 156, 74-79.

\reference{Ban17} Bannister, M., Shankman, C., Volk, K. et al. 2017, AJ, 153, 262.
  
\reference{Bat16} Batygin, K. \& Brown, M. 2016a, AJ, 151, 22.

\reference{Bat16} Batygin, K. \& Brown, M. 2016b, ApJ, 833, 3.

\reference{Bat17} Batygin, K. \& Morbidelli, A. 2017, AJ, 154, 229.

\reference{Bec17} Becker, J., Adams, F., Khain, T., Hamilton, S. \& Gerdes, D. 2017, AJ, 154, 61.

\reference{Bec18} Becker, J., Khain, T., Hamilton, S. et al. 2018, AJ, 156, 81.

\reference{Bra12} Brasser, R., Duncan, M., Levison, H., Schwamb, M. \& Brown, M. 2012, Icarus, 217, 1.

\reference{Bra15} Brasser, R. \& Schwamb, M. 2015, MNRAS, 446, 3788.

\reference{Bro04} Brown, M., Trujillo, C. \& Rabinowitz, D. 2004, ApJ, 617, 645.

\reference{Bro08} Brown, M. 2008, in The Solar System Beyond Neptune, ed. M. Barucci, H. Boehnhardt, D. Cruikshank and A. Morbidelli (Tucson: Univ of Arizona Press), 335-344.

\reference{Bro16} Brown, M. \& Batygin, K. 2016, ApJ, 824, 23.

\reference{Bro17} Brown, M. 2017, AJ, 154, 65.

\reference{Cha99} Chambers, J. 1999, MNRAS, 304, 793.

\reference{Dun08} Duncan, M., Brasser, R., Dones, L., and Levison, H. 2008, in The Solar System Beyond Neptune, eds. M. Barucci, H. Boehnhardt, D. Cruikshank and A. Morbidelli (Tucson: Univ of Arizona Press), 315-331.

\reference{Fer97} Fernandez, J. 1997, Icarus, 129, 106.

\reference{Fla15} Flaugher, B., Diehl, H., Honscheid, K., et al. 2015, AJ, 150, 150.

\reference{Gla02} Gladman, B., Holman, M., Grav, T., Kavelaars, J., Nicholson, P., Aksnes, K. \& Petit, J. 2002, Icarus, 157, 269.

\reference{Gom08} Gomes, R., Fernandez, J., Gallardo, T., \& Brunini, A. 2008, in The Solar System Beyond Neptune, ed. M. Barucci et al. (Tucson, AZ: Univ. Arizona Press) pp. 259–273

\reference{Gul10} Gulbis, A., Elliot, J., Adams, E., Benecchi, S.,
Buie, M. Trilling, D., and Wasserman, L. 2010, AJ, 140, 350-369.

\reference{Had18} Hadden, S., Li, G., Payne, M., \& Holman, M. 2018, AJ, 155, 249.

\reference{Hei86} Heisler, J. \& Tremaine, S. 1986, Icarus, 65, 13.

\reference{Hol18} Holman, M., Payne, M., Fraser, W. et al. 2018, ApJ, 855, L6.

\reference{Hol00} Holmberg, J. \& Flynn, C. 2000, MNRAS, 313, 209.

\reference{Ior17} Iorio, L. 2017, Ap\&SS, 362, 11.

\reference{Jam01} Jammalamadaka, S. \& SenSupta, A. 2001, in Topics of Circular Statistics. (World Scientific Publishing).

\reference{Kab08} Kaib, N. \& Quinn, T. 2008, Icarus, 197, 221.

\reference{Kai09} Kaib, N., Becker, A., Jones, L. et al. 2009, ApJ, 695, 268.

\reference{Kai09} Kaib, N. and Quinn, T. 2009, Science, 325, 1234.

\reference{Kai11} Kaib, N., Roskar, R., \& Quinn, T. 2011, Icarus, 215, 491.

\reference{Kai16} Kaib, N. \& Sheppard, S. 2016, AJ, 152, 133.

\reference{Ken04} Kenyon, S. \& Bromley, B. 2004, Nature, 432, 598.

\reference{Kha18} Khain, T., Batygin, K. \& Brown, M. 2018, AJ, 155, 250.

\reference{Lev01} Levison, H., Dones, L. \& Duncan, M. 2001, AJ, 121, 2253.

\reference{Li2018} Li, G., Hadden, S., Payne, M. \& Holman, M. 2018, arXiv:1806.068687.

\reference{Mad18} Madigan, A., Zderic, A., McCourt, M. \& Fleisig, J. 2018, arXiv:1805.03651.

\reference{Mal16} Malhotra, R., Volk, K. \& Wang, X. 2016, ApJ, 824, L22.

\reference{Mo04} Morbidelli, A. and Levison, H. 2004 ApJ, 128, 2564.

\reference{Mil17} Millholland, S. \& Laughlin, G. 2017, AJ, 153, 91.

\reference{Nes17} Nesvorny, D., Vokrouhlicky, D., Dones, L., Levison, H., Kaib, N. \& Morbidelli, A. 2017, ApJ, 845, 27.

\reference{Pre92} Press, W., Teukolsky, S., Vetterling, W., Flannery, B. 1992, in Numberical Recipes in C: The Art of Scientific Computing, Second Edition. (Cambride University Press) pp. 623-628.

\reference{Rab12} Rabinowitz, D., Schwamb, M., Hadjiyska, E., and Tourtellotte, S. 2012, AJ, 144, 140. 

\reference{Ric08} Rickman, H., Fouchard, M., Froeschle, C., Valsecchi, G. 2008, CeMDA, 102, 111.

\reference{Sef18} Sefilian, A. \& Touma, J. 2018, arXiv:1804.06859.

\reference{Sha17} Shankman, C., Kavelaars, J., Lawler, S., Gladman, B. \& Bannister, M. 2017a, AJ, 153, 63.

\reference{Sha17} Shankman, C., Kavelars, J., Bannister, M. et al. 2017b, AJ, 154, 50.

\reference{She11} Sheppard, S., Udalski, A., Trujillo, C. et al. 2011, AJ, 142, 98.

\reference{She16} Sheppard, S., Trujillo, C., \& Tholen, D. 2016, ApJ, 825, L13.

\reference{She16} Sheppard, S. \& Trujillo, C. 2016, AJ, 152, 221.

\reference{Soa13} Soares, J., and Gomes, R. 2013, AA, 553, 110.

\reference{Tru14} Trujillo, C. \& Sheppard, S. 2014, Nature, 507, 471.

\reference{Tru18} Trujillo, C. 2019, ApJ, submitted.

\end{references}
\end{document}